\documentclass[aps,10pt,prc,reprint,eqsecnum,nofootinbib]{revtex4-2}
\usepackage{graphicx}
\usepackage[tbtags]{amsmath}
\usepackage{bm,amssymb}

\newcommand{\br}{{\bm{r}}}
\newcommand{\bp}{{\bm{p}}}
\newcommand{\bq}{{\bm{q}}}

\newcommand{\Tr}{\mathop{\rm Tr}}
\newcommand{\eps}{\varepsilon}

\def\<{\langle}
\def\>{\rangle}
\def\NN{\nonumber \\}
\usepackage{mathptmx}

\begin{document}
\title{Semiclassical description of quadrupole-hexadecapole correlation
in nuclear shape evolution}
\author{Ken-ichiro Arita\email{arita@nitech.ac.jp}}
\affiliation{Department of Engineering Physics, Nagoya Institute of Technology,
Nagoya 466-8555, Japan}
\date{\today}

\begin{abstract}
\noindent
\textbf{Background:}
Theoretical investigations of nuclear shapes with realistic effective
interactions or mean field potentials have suggested a remarkable
systematics in the hexadecapole shape evolution with
varying particle number: Within each single-particle shell, from one
spherical magic number to the next, a diamond type ($\beta_4>0$)
appears in the first half, and then turn into oblong type
($\beta_4<0$) in the second half. \\
\textbf{Purpose:}~
Nuclear deformation is essentially governed by the single-particle
shell structure.  In semiclassical periodic orbit theory (POT), level
density is expressed as the sum over contributions from classical
periodic orbits, and the origins of the gross shell structures can be
understood by the contribution of one or a few shortest orbits.  Using
the POT, I examine how the hexadecapole degree of freedom affect the
contribution of the orbits to the deformed shell structure through
their bifurcations to explain the mechanism of above shape
evolution. \\
\textbf{Methods:}~
Ground state deformations are systematically investigated by the
shell correction method, taking
account of axially symmetric quadrupole and hexadecapole shape degrees
of freedom.  For simplicity, I employ two simplest mean field
potentials to obtain the shell corrections: the oscillator and the
cavity (infinite well) potentials.  After confirming the
the systematics in shape evolution under these simplest
mean-field models, semiclassical analyses are made, focusing on the
role of PO bifurcation. \\
\textbf{Results and conclusions:}~
Systematics in the hexadecapole shape evolution in each
single-particle shell is clearly explained by the bifurcations of
two different types of periodic orbits in the cavity model, which
suggest strong
correlation between quadrupole and hexadecapole parameters.
This also ensures the mechanism of nuclear prolate-shape predominance
within this simple potential model.
\end{abstract}
\maketitle

\section{Introduction}

Nuclei take variety of shapes depending on the numbers of protons and
neutrons, and distinct regularities are observed in the way
for deformations to manifest.
The most significant aspect is the periodic appearances of strongly bound
spherical ``magic'' states corresponding to the
closed-shell configurations of the single-particle energy levels in the
spherical mean field potential.  Nuclei
with nucleon numbers outside those magic numbers are generally
deformed, where the quadrupole shape degrees of freedom play
leading roles.  For medium-mass to heavy nuclei, the
mean-field potential have sharp surface (in comparison to the harmonic
oscillator model) and it causes asymmetry
in the single-particle shell structure for prolate and oblate sides.
It has long been known that most of the ground-state deformations of
nuclei are of the prolate type\cite{BMText2}, and the above
prolate-oblate asymmetry of the shell structure is considered one of
the primary reasons of the prolate shape dominance.  Various theoretical
approaches have been attempted to elucidate the origin of
this asymmetry\cite{HamMot09,Tajima02,Takahara12,Hor10},
including those employing semiclassical
periodic-orbit theory\cite{Frisk90,Arita12,Arita16}.
In those analyses, hexadecapole shape degree of freedom has not
received explicit attention.

In the multipole expansion of the surface shape,
octupole degrees of freedom follow the quadrupole ones and these
break the reflection symmetry.  It has been suggested that a few
isotopes locating ``northeast'' neighbors of doubly-magic
configurations on the nuclear chart $(N,Z)$ break reflection symmetry
in their ground states\cite{Butler96,HamMot91,Rob11,Mol16,Ebata17,Cao20}.
Even if they do not exhibit static octupole deformation, strong
octupole correlation is suggested for isotopes in these regions.
The octupole deformation (correlation) for those nuclei have been
attributed to the existence of nearly degenerate $\Delta l=3$
single-particle orbitals lying just above the energy gap\cite{Butler96}.
In addition, I have pointed out the role of the gross shell effect
associated with the local restoration of dynamical
symmetry\cite{Arita23a,Arita23b}.
Non-axial octupole degrees of freedom, such as tetrahedral shape, have
also been examined in various theoretical
approaches\cite{HamMot91,Schunck04,Tagami13,AriMuk14}.

Hexadecapole shape degrees of freedom also play significant
roles in lowering shell energy.
Recently, the interplay between quadrupole and
hexadecapole shape degrees of freedom were discussed in
multi-reference GCM calculations\cite{KumRob23}, which elucidated the
significance of hexadecapole degree of freedom.
Theoretical calculations of
nuclear mass throughout the nuclear chart with modern realistic
mean fields or effective interactions\cite{Mol16,KumRob23} have prevailed
a remarkable systematics in ground-state hexadecapole deformation.
In each single-particle shell from one spherical magic to the next,
diamond type shapes ($\beta_4>0$) appear
in the first half and then turn into oblong type ($\beta_4<0$) in
the second half.
This systematics has also been suggested by a simple consideration of
single-particle
orbitals filling the spherical shells\cite{Suekane57,Ber67}.
In experiments, identification of hexadecapole deformation
is generally a difficult task, as it is buried in quadrupole deformation.
Recently, new experimental techniques to access hexadecapole moment
have been proposed such as backward quasi-elastic
scattering\cite{Jia2014} and heavy ion collision\cite{Xu2024}, and
more information on hexadecapole deformation for nuclei
might be accessible in near future.

In my recent work on the systematics of ground-state octupole
deformations using the semiclassical POT, bifurcations of the POs,
associated with local dynamical symmetry restoration, play a
significant role in stabilizing shapes with specific combinations of
quadrupole and octupole parameters.  In the present work, I will
verify whether the above semiclassical mechanism for the enhancement
of deformed shell effect also provides an explanation of the
systematics in hexadecapole shape evolution.  To this aim, I employ
two simplest mean-field models, oscillator and cavity, as in my
previous work on octupole deformation\cite{Arita23b}.  In
Sec.~\ref{sec:quantum}, the mean-field models used in this calculation
are defined.  Using these models, which take quadrupole and
hexadecapole shape degrees of freedom into account, ground state
shapes are calculated for each system with $N$ neutrons and $Z$ protons
across the entire nuclear chart.  In
Sec.~\ref{sec:theory}, semiclassical theory of gross shell structure
is briefly summarized, mainly focusing the role of classical PO
bifurcations.  Then, semiclassical analysis is presented in
Sec.~\ref{sec:calc}
for cavity model which qualitatively reproduces the systematics
of shape evolution in a realistic mean field model.
Section~\ref{sec:summary} is devoted to summary
and discussion.

\section{Systematics of quadrupole-hexadecapole shapes in simple
potential models.}
\label{sec:quantum}

\subsection{The model}

There are several ways of parametrizing shape of nuclear surface.  The
most popular one is a simple multipole expansion which has been
usually adopted to the calculations with the Woods-Saxon type
potential models.  In this work, I employ the stretched-multipole
shape parametrization as in my previous work on the octupole
deformation\cite{Arita23b}.
\begin{figure}[t]
\includegraphics[width=.75\linewidth]{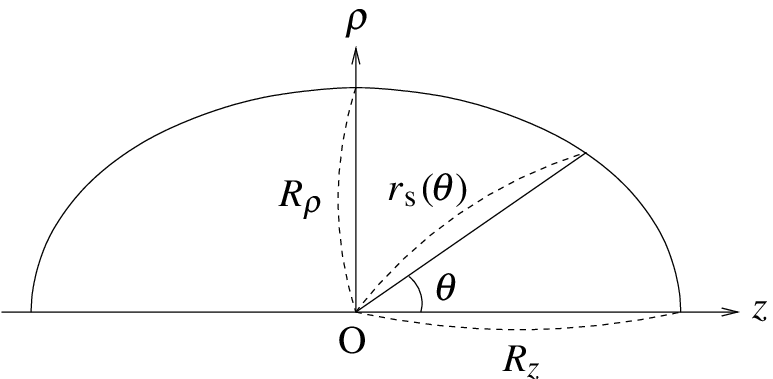} 
\caption{\label{fig:profile}
Nuclear surface profile function $r_{\rm s}(\theta)$ for axially
symmetric deformation.  An example for a spheroidal shape.}
\end{figure}
Let $r_s(\theta)$ be the distance from
the origin O to the potential surface in the direction of the polar
angle $\theta$ as shown in Fig.~\ref{fig:profile}.
For axially symmetric deformation, $r_s(\theta)$ is given by
\begin{equation}
r_s(\theta;\beta)=R_0 f(\cos\theta;\beta),
\end{equation}
where $\beta$ is the deformation parameter (or set of the
parameters) and $R_0$ is the
radius of the surface in the spherical limit.  Due to the volume
conservation condition, the shape function $f$ satisfies
\begin{equation}
\frac12\int_{-1}^1 f^3(u;\beta)du=1, \quad f(u;0)=1.
\end{equation}
To describe the quadrupole type deformation, I employ the spheroidal
shape parametrization.  The shape function for pure
spheroidal shape with axis ratio $\eta=R_z/R_\rho$
(see Fig.~\ref{fig:profile}) is give by
\begin{gather}
f_2(u;\eta)=\frac{\eta^{2/3}}{\sqrt{u^2+\eta^2(1-u^2)}}.
\end{gather}
When the hexadecapole deformation is considered, it is first applied
on the spherical surface $(\eta=1)$ with the parameter $\beta_4$, and
then stretched (or shrunk) in direction along the symmetry axis to
combine with quadrupole deformation.  In this way, the surface shape
is parametrized by $\beta_4$ and $\eta$ as
\begin{gather}
r_s(\theta)=R_0 f_2(u;\eta)e^{\beta_4 P_4(u')}, \NN
u=\cos\theta, \quad u'=\frac{u}{\sqrt{\eta^2-(\eta^2-1)u^2}}.
\end{gather}
By putting $P_4$ function in the exponent, a natural shape profile is
provided up to large $\beta_4$ in comparison to a conventional linear
dependence $r_s=R_0(1+\beta_4P_4)$.
Dimensionless quadrupole and stretched-hexadecapole moments are
defined as
\begin{gather}
q_2=\frac{1}{R_0^2}\<r^2 P_2(\cos\theta)\>, \\
q_4=\frac{1}{R_0^4}\<(r^4 P_4(\cos\theta))'\>
   =\frac{1}{R_0^4}\<r^4P_4(\cos\theta)\>_{\eta\to1},
\end{gather}
where the average $\<*\>$ is calculated with an assumed single-nucleon
density proportional to the potential depth from the Fermi level.
The useful feature of this parametrization is that the quadrupole and
hexadecapole parameters are mutually independent.  Change of
hexadecapole parameter $\beta_4$ (``quadrupole'' parameter $\eta$)
does not affect the $q_2$ ($q_4$), which is not the case with the
simple multipole expansion.  The surface shape for $q_2=0.1$
$(\eta\approx 1.2)$ with several values of $q_4$ are displayed in
Fig.~\ref{fig:shapes}.
\begin{figure}[t]
\includegraphics[width=\linewidth]{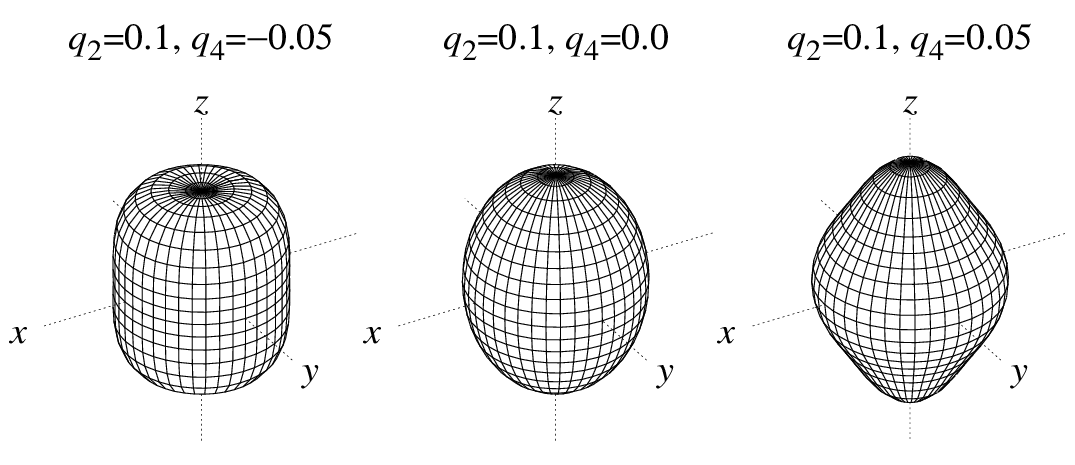} 
\caption{\label{fig:shapes}
Surface shape for combinations of quadrupole and hexadecapole
deformations.}
\end{figure}

Using the above shape parametrization, single-particle potential
having the power-law type radial dependence is given by
\begin{equation}
U(\br)=U_0\left(\frac{r}{r_s(\theta;\beta)}\right)^\alpha,
\quad \beta=\{q_2,q_4\}.
\label{eq:mpot}
\end{equation}
With this choice of the potential, the single-particle Hamiltonian
follows the scaling law:
\begin{equation}
H(c^{1/2}\bp,c^{1/\alpha}\br)=cH(\bp,\br), \label{eq:scaling}
\end{equation}
which extremely simplifies the semiclassical analysis\cite{Arita12}.
By a suitable choice of the power parameter $\alpha$, this potential
provides a good approximation to the Woods-Saxon potential over a wide
range of mass numbers.  With inclusion of spin-orbit term (which is
not considered in the current work), it can be even employed as
semi-realistic mean field model for medium-mass to heavy
nuclei\cite{Arita16}.  For numedical calculations in the following
part, I will consider only two limiting cases, oscillator ($\alpha=2$)
and cavity ($\alpha=\infty$), because no good spherical magic numbers
are obtained for intermediate values of $\alpha$ without spin-orbit
coupling.

\subsection{Ground state hexadecapole deformations}

\begin{figure}
\centering
\includegraphics[width=\linewidth]{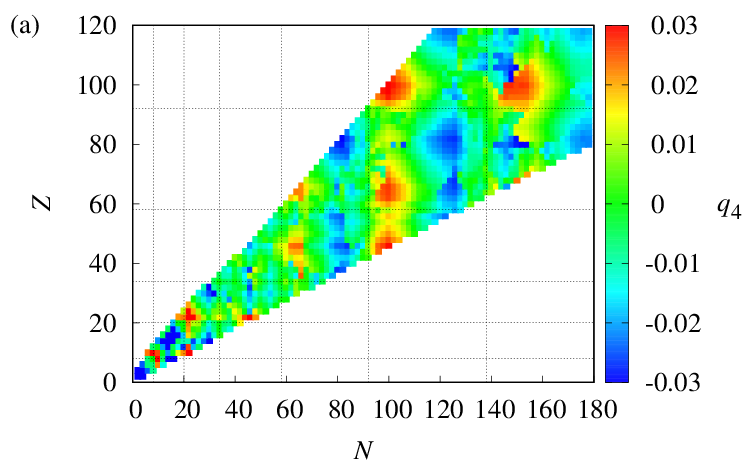} \\ 
\includegraphics[width=\linewidth]{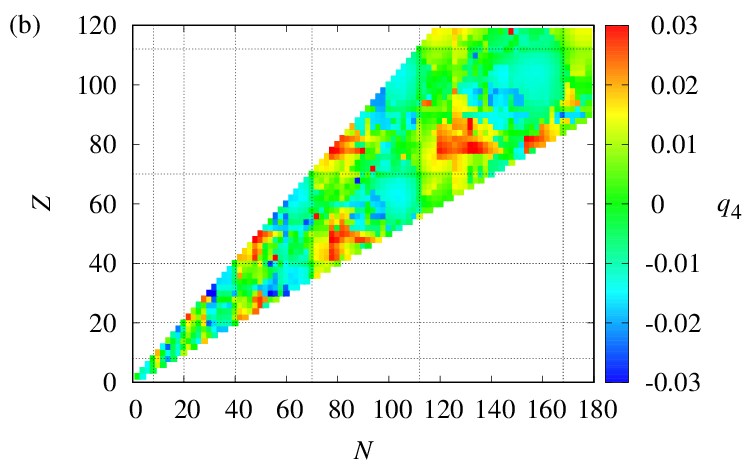} \\ 
\caption{\label{fig:echart}
Ground-state hexadecapole deformation calculated 
for nuclei with $Z\leq N\leq 2Z$ by the shell
correction method with (a) cavity and (b) oscillator type potential
models.  Horizontal and vertical dotted lines indicate the
spherical magic numbers in each model.}
\end{figure}

Using the shell correction method with the above mean-field models,
I first calculate ground-state energies of the
systems with $N$ neutrons and $Z$ protons as functions of shape
parameters.  Using the single-particle
energies $\{\eps_i\}$ obtained for the Hamiltonian $h=t+u$ which
consists of kinetic energy $t$ and mean-field potential $u$ with given
deformation, single-particle energy sum $E_{\rm s.p.}$ is calculated,
and then it is decomposed into the smooth part $\tilde{E}$ and the
oscillating part $\delta E$ as
\begin{equation}
E_{\rm s.p.}(N)=\sum_{i=1}^N\eps_i=\tilde{E}_{\rm s.p.}(N)+\delta E(N).
\end{equation}
The above smooth part will be given by the sum of the
averaged values of kinetic energy and mean field potential as
$\tilde{E}_{\rm s.p.}=\<t\>+\<u\>$.  Assuming $u$ as what is deduced from a
two-body interaction, the smooth part of the many-body energy should
be given by $\tilde{E}=\<t\>+\frac12\<u\>$ to avoid the double
counting of the interaction.  Using the Virial-theorem relation
for our power-law potential model
$\frac12\<t\>=\frac{1}{\alpha}\<u\>$, one obtains\cite{Arita12}
\begin{equation}
\tilde{E}(N,Z)=\frac{\alpha+1}{\alpha+2}\left(\tilde{E}_{\rm s.p.}(N)
 +\tilde{E}_{\rm s.p.}(Z)\right).
\end{equation}
For the cavity potential $(\alpha=\infty)$, the above prescription
gives $\tilde{E}=\tilde{E}_{\rm s.p.}$.  However, it is known that
this relation significantly underestimates the relative magnitude of
the shell effect.  Here, in somewhat an ad hoc way, I set
$\tilde{E}=\frac12\tilde{E}_{\rm s.p.}$ which might be acceptable
since qualitative results are not very sensitive to this choice.
Figure~\ref{fig:echart} shows the systematics of the ground-state
hexadecapole deformation on the nuclear chart.  The upper panel (a) is
for the cavity (infinite-well, $\alpha=\infty$) potential model, and
the lower panel (b) is for the oscillator-type $(\alpha=2)$ potential
model.  Vertical and horizontal dotted lines indicate spherical magic
numbers for each potential model: $N(Z)=8,20,34,58,92,138,\cdots$ for
the cavity and $8,20,40,70,112,168,\cdots$ for the oscillator.  In the
both panels, positive hexadecapole deformation systematically appear
at the `northeast' neighbors of the doubly-magic nuclei (corresponding
to the crossing
points of the vertical and horizontal dotted lines) and then it turns
negative as approaching the `southwest' neighbors of doubly magic one.
This suggests that
these simple models already include the essential mechanism for the
systematics in the ground-state hexadecapole shape evolution found in
the realistic models.

Deformation is essentially governed by the gross shell
structure.  In the following, I will examine the shell effect
responsible for the above systematic hexadecapole shape evolution from
a spherical magic to the next, and clarify its semiclassical origin
using the periodic orbit theory.

\section{Semiclassical theory of shell structure}
\label{sec:theory}

\subsection{Trace formula for level density and shell energy}

In general, density of energy eigenvalues for a particle confined in a
three-dimensional potential shows a gross oscillating pattern
which is called a shell effect.
Except for several integrable systems such as Coulomb and harmonic
oscillator potential models, the origin of such fluctuation cannot be
explained by a purely quantum mechanical concept.  In early 1970s,
semiclassical formulas were found which express quantum level density
in the form of the summation over contributions of classical periodic
orbits (POs) \cite{Gutz71,BaBlo72,GutzText,BBText}.  Those formulas, which
can be only applied to a limited class of systems in their original
forms, have been extended to those applicable to more generic
systems\cite{BerTab76,StrMag76,CreLJ91,Cre93,Sie96,SchSie97,Sie98}.
The general form of the semiclassical level density is expressed as
\begin{align}
g(e)&=\sum_n\delta(e-e_n) \NN
&=\bar{g}(e)+\sum_{\rm po}A_{\rm po}(e)
\cos\left(\tfrac{1}{\hbar}S_{\rm po}(e)-\tfrac{\pi}{2}
\mu_{\rm po}\right). \label{eq:trace_g}
\end{align}
$S_{\rm po}=\oint_{\rm po}\bp\cdot d\br$ is the action integral along
the PO and it is generally a monotonically increasing function of
energy.  $\mu_{\rm po}$ is the Maslov
index which is related to the geometric character of the orbit.  It
should be emphasized that the contribution of each PO is not
related to individual quantum levels but to a certain
periodic structure embedded in the distribution of the levels.

Derivation of this formula is based on the stationary-phase method
for evaluating integrals including the path integral.
Let us consider the path integral representation of the level density
\begin{align}
g(e)&=\Tr\delta(e-\hat{h}) \NN
&=\frac{1}{2\pi i\hbar}\int_{-\infty}^{\infty}dt\int d\br
\<\br|e^{i(e-\hat{h})t/\hbar}|\br\> \NN
&=\frac{1}{2\pi i\hbar}\int_{-\infty}^\infty dt\,e^{iet/\hbar}
\int d\br\int D[\br_q]e^{iR[\br_q(\tau)]/\hbar},
\end{align}
where $R[\br_q(\tau)]$ represents the action integral along the closed
path $\br_q(\tau)$ $(0\leq\tau\leq t)$ which starts from
$\br_q(0)=\br$ at time $\tau=0$ and returns to $\br_q(t)=\br$ at
$\tau=t$,
\[
R[\br_q(\tau)]=\int_0^t d\tau\,L(\br_q(\tau),\dot{\br}_q(\tau)).
\]
$L(\br,\dot{\br})$ is the Lagrangian
and $D[\br_q]$ represents the integration measure of the path $\br_q$
in the path integral.  Noting that the stationary phase condition with
respect to the variation of the path is equivalent to the Hamilton's
principle which derives the classical equations of motion, the
resulting formula appears as a sum over the terms associated with the
classical closed trajectory.  Evaluation of the trace integral by the
stationary-phase method extracts the periodic orbit because the
stationary-phase condition requires the matching of initial and final
momenta of the closed trajectory.  In these procedures, continuous
symmetry of the system should be properly taken into account which
generally leads to a degeneracy of the periodic orbits.  For a $K$
parameter family of orbit, $K$ integrals should be carried out exactly
and other integrals are evaluated by the saddle-point approximation as
\begin{gather}
\int d^fx\, e^{iS(x)/\hbar}
\simeq\sum_i\sqrt{\frac{(2\pi i\hbar)^f}{\det S''(\bar{x}_i)}}
e^{iS(\bar{x}_i)/\hbar}
\label{eq:spa}
\end{gather}
where $\bar{x}_i$ are the stationary points of the action satisfying
$(\partial S/\partial x)_{\bar{x}_i}=0$.  Since the dimension $f$ of
the integration for a degenerate orbit is smaller by $K$ than that
for an isolated orbit, one finds the semiclassical order of the
contribution of degenerate POs forming a $K_{\rm po}$-parameter family
to be $A_{\rm po}\propto \hbar^{-K_{\rm po}/2}$.

\begin{figure}[t]
\includegraphics[width=\linewidth]{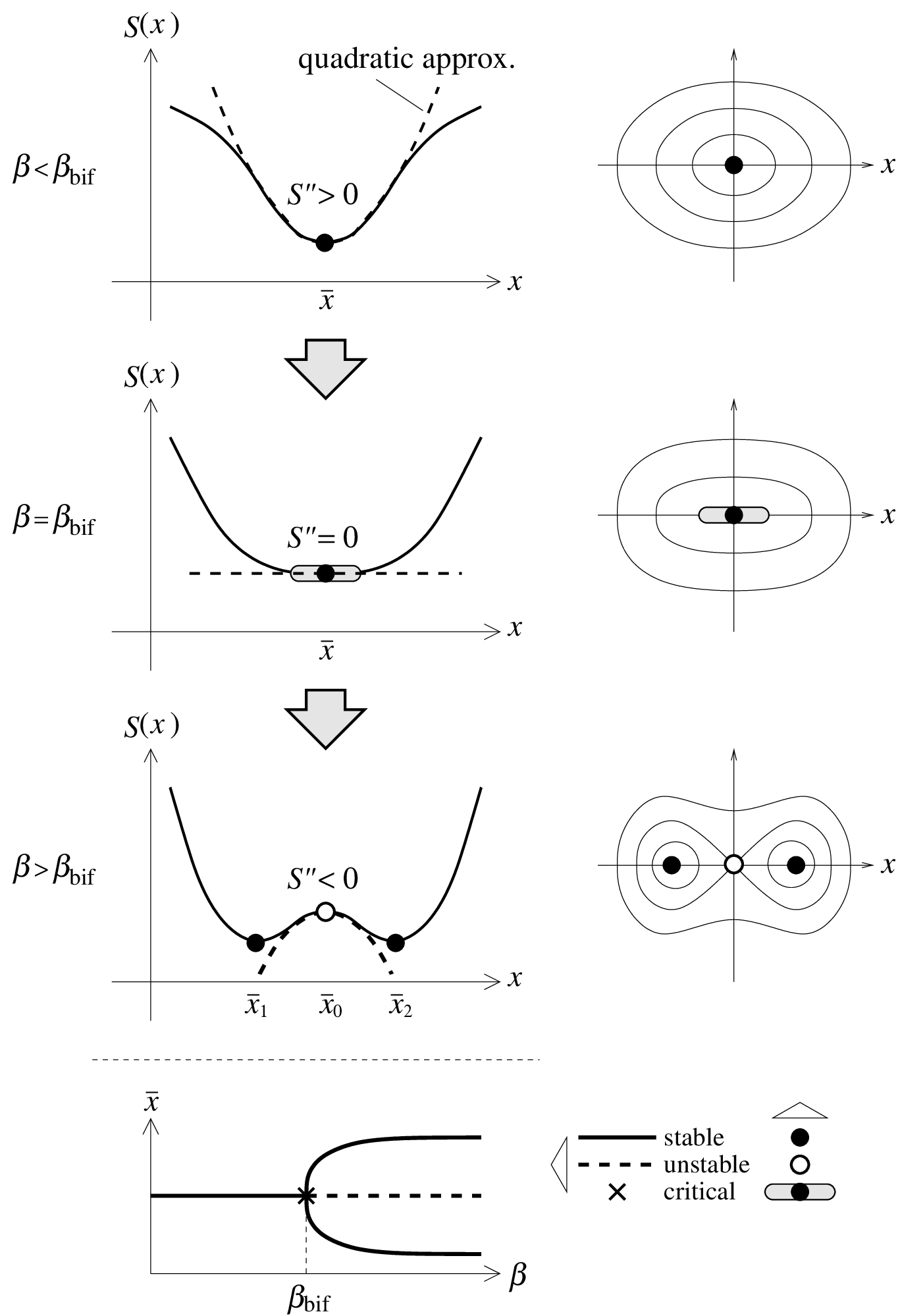} 
\caption{\label{fig:bif_po}
Illustration of the periodic orbit bifurcation in case of the
pitchfork type.  Upper three panels describe the intersections of
action $S$ along the relevant coordinate $x$ (the left-hand side),
and the behavior of the Poincar\'{e} surface (the right-hand side),
with varying deformation parameter $\beta$ around the bifurcation
point $\beta_{\rm bif}$.
The bottom panel illustrates the behaviors of stationary points
$\bar{x}$ as functions of $\beta$.}
\end{figure}

Here, let us focus on the integral with respect to a certain
coordinate $x$ which is relevant to a bifurcation.
Each stationary point corresponds to a periodic orbit,
and the curvature $S''$ of the action at the stationary point
describes the stability of the orbit against the deviation of the
initial point to $x$ direction.
Here, stability refers to how a trajectory diverges from the
original one when initial conditions are perturbed.  Stable orbits
stay around the original one, but unstable
orbits diverge exponentially as functions of time.
With varying deformation parameter $\beta$, the curvature $S''$ may
change its sign at a certain value $\beta=\beta_{\rm bif}$.
This sign reversal is generally accompanied by a bifurcation of PO
because the number of stationary points changes.  Fig.~\ref{fig:bif_po}
illustrates it for a bifurcation of ``pitchfork'' type.
In this bifurcation, the mother PO which is stable at
$\beta<\beta_{\rm bif}$ turns unstable at $\beta>\beta_{\rm bif}$ and
two stable daughter POs emerge on both sides of the mother.
Near the bifurcation point $\beta_{\rm bif}$, a family of
quasi-periodic orbits are generated around the PO
and it makes coherent contribution to the integral and enhance the
amplitude $A_{\rm po}$ in the PO sum (\ref{eq:trace_g}).  Such
enhancement is especially remarkable in the case of short and simple
POs\cite{AriBra08b},
and an enhancement of gross shell effect can be expected around their
bifurcation points.  It plays a crucial role in the formation of
a prominent shell structure under specific deformation.  At the
bifurcation point where the curvature $S''$ at the stationary point
vanishes, the simple saddle point approximation (\ref{eq:spa}) breaks
down, and a higher order expansion around the stationary point should
be considered to evaluate the amplitude factor, e.g., by the uniform
approximations\cite{Sie96,Sie98}.

By using the POT trace formula (\ref{eq:trace_g}) for the level density,
one can obtain the semiclassical PO-sum formula for shell
energy.  According to the Strutinsky's theorem of shell
correction\cite{StrMag76,BBText},
shell energy is expressed in terms of the fluctuation
part of the level density as  
\begin{equation}
\delta E(N)=\int_0^{e_F}(e-e_F)\delta g(e)de. \label{eq:sce_dg}
\end{equation}
$e_F$ can be taken as the Fermi energy associated with the averaged
level density $\bar{g}(e)$, and is determined as a function of
particle number $N$ by
\begin{equation}
\int_0^{e_F}\bar{g}(e)de=N.
\end{equation}
Inserting the semiclassical level density (\ref{eq:trace_g}) in
Eq.~(\ref{eq:sce_dg}), the trace formula for the shell energy is
obtained %
\footnote{The derivation is based on the evaluation of the integral
over energy $e$ by the partial integration and extracting the end-point
contribution at $e\simeq e_F$ in the integration of rapidly oscillating
integrand\cite{BBText}.}
as
\begin{equation}
\delta E(N)=\sum_{\rm po}\frac{A_{\rm po}(e_F)}{(T_{\rm po}/\hbar)^2}
\cos\left(\tfrac{1}{\hbar}S_{\rm po}(e_F)-\tfrac{\pi}{2}\mu_{\rm po}\right).
\label{eq:trace_sce}
\end{equation}
In this formula, the contributions of longer orbits are strongly
suppressed by the additional factor $1/T_{\rm po}^2$, where
$T_{\rm po}(e)=dS_{\rm po}(e)/de$ represents the period of the orbit.
Consequently, the
shell energy is dominated by the contributions of a few
shortest POs.
In this way, POT is particularly useful in the analysis of shell energy.

\section{Semiclassical analysis of the hexadecapole shape evolution}
\label{sec:calc}

\subsection{Fourier analysis}

Since our Hamiltonian has the scaling rule (\ref{eq:scaling}),
one has the same set of POs independently of energy $e$ for a given shape
$\beta$, and the action integral can be factorized as
\begin{gather}
S_{\rm po}(e,\beta)=\eps\tau_{\rm po}(\beta), \\
\tau_{\rm po}(\beta)=S_{\rm po}(e_0,\beta), \quad
\eps=\left(\frac{e}{e_0}\right)^{1/2+1/\alpha},
\end{gather}
where $e_0$ is an appropriate energy unit.
For a cavity system ($\alpha=\infty$), $\eps$ can be taken as wave
number $k$ and then $\tau_{\rm po}$ becomes the orbit length $L_{\rm po}$.
With the wave number variable, trace formula (\ref{eq:trace_g}) is
rewritten as
\begin{align}
g(k)&=\frac{de}{dk}g(e) \nonumber \\
&=g_0(k)+\sum_{\rm po}A_{\rm po}(k)
\cos\left(kL_{\rm po}-\tfrac{\pi}{2}\mu_{\rm po}\right).
\label{eq:trace_cavity}
\end{align}
$g_0$ represents the average level density
corresponding to the Thomas-Fermi (or the extended Thomas-Fermi)
approximation.
For this scaling system, Fourier transformation technique provides
a convenient way of examining quantum-classical
correspondence\cite{BaBlo72,Arita12}.
Consider the Fourier transform of the level density
\[
F(L)=\int e^{ikL}g(k)dk.
\]
Inserting the quantum level density $g(k)=\sum_i\delta(k-k_i)$,
it is simply written as $F(L)=\sum_i e^{ik_i L}$.
Practically, summation over infinite number of eigenvalues in the
above expression is impossible, and a gaussian cutoff
factor is inserted in the integrand as
\begin{equation}
F(L)=\sqrt{\frac{2}{\pi}}\frac{1}{k_c}
\int_0^\infty e^{-\frac12(k/k_c)^2}g(k)dk. \label{eq:fourier}
\end{equation}
Then, it allows us to evaluate $F(L)$ with a finite number of quantum levels
$k_i$ that do not significantly exceed $k_c$,
\begin{equation}
F_{\rm qm}(L)=\sqrt{\frac{2}{\pi}}\frac{1}{k_c}
\sum_i e^{-\frac12(k_i/k_c)^2} e^{ik_iL}. \label{eq:fourier_qm}
\end{equation}
By inserting the semiclassical trace formula (\ref{eq:trace_cavity})
into (\ref{eq:fourier}), it turns out to be a function exhibiting
successive peaks at the lengths of POs:
\begin{gather}
F_{\rm cl}(L)=F_0(L)+\sum_{\rm po}A_{\rm po}
e^{i\pi\mu_{\rm po}/2}e^{-\frac12 k_c^2(L-L_{\rm po})^2}.
\label{eq:fourier_cl}
\end{gather}
Taking larger $k_c$ makes the width of those peaks
$\Delta L\sim\frac{1}{k_c}$ narrower, and
improves the resolution of the lengths.
To obtain a simple Gaussian expression in Eq.~(\ref{eq:fourier_cl}),
the amplitude factor $A_{\rm po}$ is assumed to be a constant.
Even when the $k$ dependence of the amplitude is correctly
taken into account, however, the Fourier transform remains a peaked
function similar
to a Gaussian (see Fig.~11 of Ref.~\cite{Arita18}).
Therefore, the Fourier transform (\ref{eq:fourier_qm}) evaluated by
the quantum mechanically calculated levels will show a series of peaks
at the lengths of classical POs with intensities corresponding to the
magnitude of amplitude factor $A_{\rm po}$.  In this way,
one can examine which
POs make critical contribution to the quantum shell structures
without directly evaluating the trace formula.

\begin{figure}
\includegraphics[width=.9\linewidth]{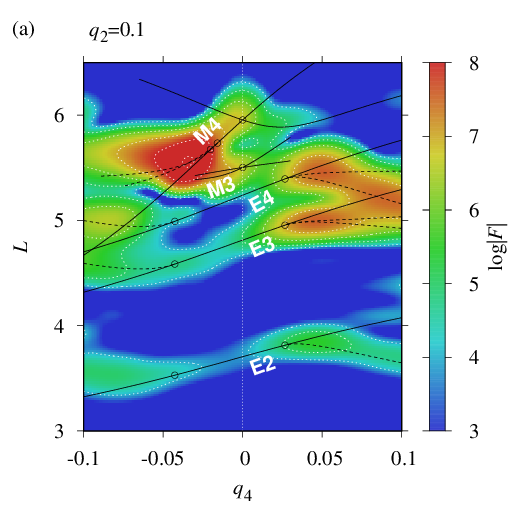} \\ 
\includegraphics[width=.9\linewidth]{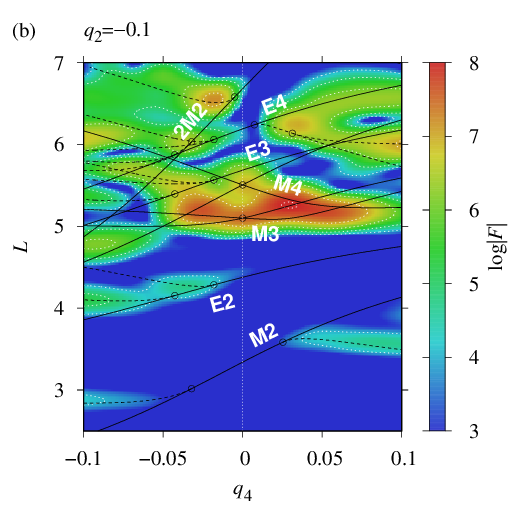}    
\caption{\label{fig:fmap}
Modulus of the Fourier transform (\ref{eq:fourier}) of the
single-particle level density as function of length parameter $L$ and
hexadecapole parameter $q_4$, with the quadrupole parameter fixed at
(a) $q_2=0.1$ and (b) $q_2=-0.1$.  Black curves represent the lengths
of the classical PO families as functions of $q_4$.
Solid lines show the lengths of the equatorial (E2--4) and meridian
(M2--4) orbits that continuously exist for all $q_4$ value.
They encounter bifurcations at the deformations indicated by the
open circle symbols and new orbits bifurcate off from them whose
lengths are shown by broken lines.}
\end{figure}

\begin{figure}[t]
\includegraphics[bb=0 35 430 210,width=.8\linewidth]{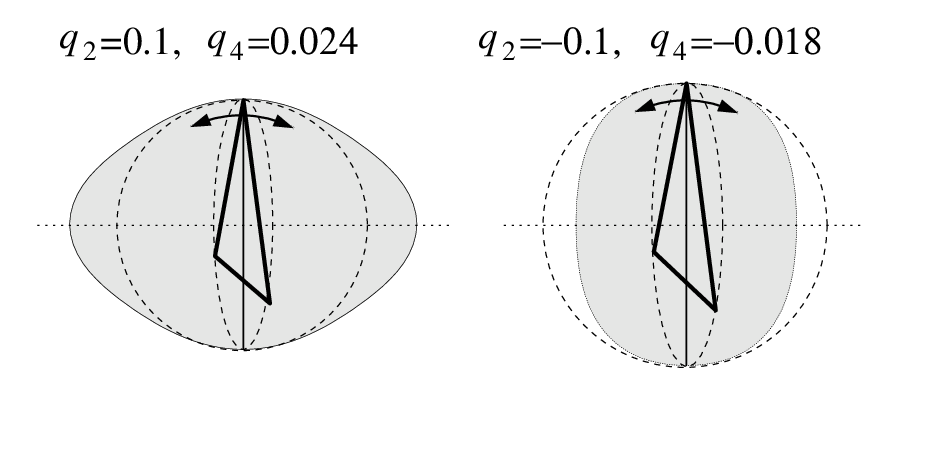} 
\caption{\label{fig:bif_sph}
Restoration of local spherical symmetry for certain combinations of
quadrupole and hexadecapole deformations.  Horizontal dotted line
represents the symmetry axis and the vertical solid line represents
the equatorial diameter orbit.  Circle with broken line indicates the
sphere whose radius is equal to the curvature radius of the surface
at the equatorial plane.}
\end{figure}

\begin{figure}[t]
\includegraphics[bb=0 45 430 195,width=.8\linewidth]{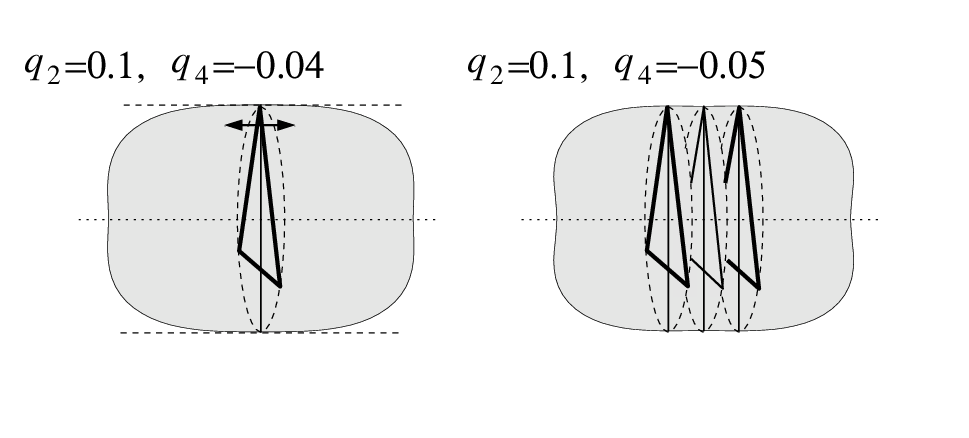} 
\caption{\label{fig:bif_asym}
Bifurcation of equatorial orbits with fixed $q_2=0.1$ and decreasing
$q_4$.  The curvature radius of the surface at the equatorial plane
changes sign at $q_4\sim -0.04$, which generates additional orbital
planes parallel to the equatorial plane on both sides of it.}
\end{figure}

\begin{figure}
\includegraphics[width=\linewidth]{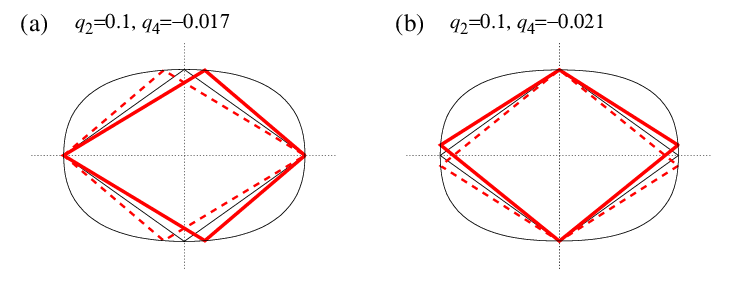} \\ 
\includegraphics[width=\linewidth]{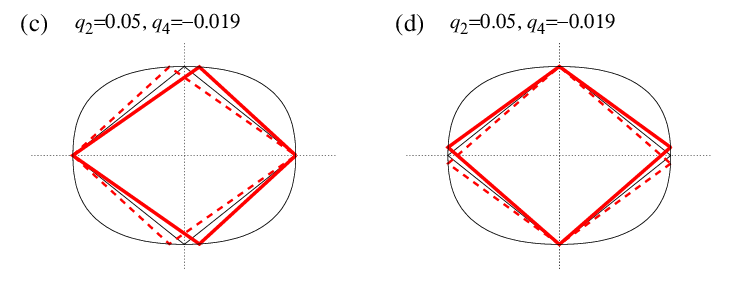}    
\caption{\label{fig:traj_md}
Meridian-plane diamond orbit and the daughter orbits bifurcated off
from it.  Thick solid and broken lines in each panel represent
the daughters that are mirror images of each other and have identical
properties.}
\end{figure}

Figure~\ref{fig:fmap} shows the correspondence between peaks in
modulus of the Fourier
transform (\ref{eq:fourier_qm}), evaluated by the quantum levels, and the
lengths of classicall POs.  In this calculation, quadrupole parameter is
fixed at $q_2=0.1$ in the upper panel (a) and $q_2=-0.1$ in the lower
panel (b), and
the hexadecapole parameter $q_4$ is varied from $-0.1$ to $0.1$.  At
$q_4=0$, meridian-plane triangular (M3) and parallelogram (M4) family
of orbits play significant role because of the extra degeneracy
associated with the dynamical symmetry of spheroidal cavity system.
Let us first discuss the the prolate case in the upper panel (a) of
Fig.~\ref{fig:fmap}.  With increasing $q_4>0$, curvature
of the potential surface at the equatorial plane becomes larger and the
curvature radius coincides with the equatorial radius at
$q_4\sim 0.024$, and then the
spherical symmetry is locally restored in vicinity of the equatorial plane
(see the left panel of Fig.~\ref{fig:bif_sph}).
This local symmetry
restoration causes bifurcations of the equatorial POs and new families
of orbits emerge, that are not on the equatorial plane but are rotated
as indicated by the arrows.  One can see
enhancement of the Fourier peaks corresponding to the equatorial
diametric (E2), triangular (E3) and square (E4) orbits around
this bifurcation point.  With decreasing $q_4<0$, curvature
at the equatorial plane decreases and becomes zero at $q_4\sim -0.04$.
Thereafter, asymmetric orbit planes parallel to the equatorial plane
are generated as in the right panel of Fig.~\ref{fig:bif_asym},
and new POs on them emerge via the bifurcations of the
equatorial ones.
One finds some enhancement in the Fourier amplitudes of the equatorial
orbits there, but they are not as remarkable as in the
$q_4>0$ side.  This may be related to the difference in the
degeneracy of the local symmetry: $K=3$ for a sphere
and $K=2$ for a cylinder.
One also notices a significant enhancement of the Fourier amplitude
for meridian-plane diamond orbit M4, which encounters bifurcations
around $q_4\simeq-0.02$ (see Fig.~\ref{fig:traj_md}).
For $q_2=0.1$, this orbit encounters bifurcations of two types shown in
the panels (a) and (b) at different values of $q_4$.
Furthermore, if I also change the value of $q_2$, coalescence of these two
bifurcation points is found at $(q_2,q_4)\simeq (0.05,-0.019)$ as shown in
panels (c) and (d).
Here, two different pitchfork bifurcations occur almost
simultaneously, which leads to a
local symmetry of higher dimension.  This kind of bifurcation
interference is called a codimension-two
bifurcation\cite{Schom97,AriBra08b}.
This peculiar bifurcation is expected to play significant roles
in deformed shell structure.

\begin{figure*}
\begin{flushright}
\includegraphics[height=75mm]{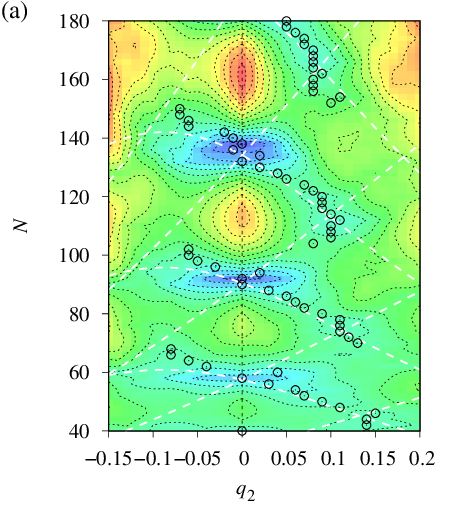} 
\includegraphics[height=75mm]{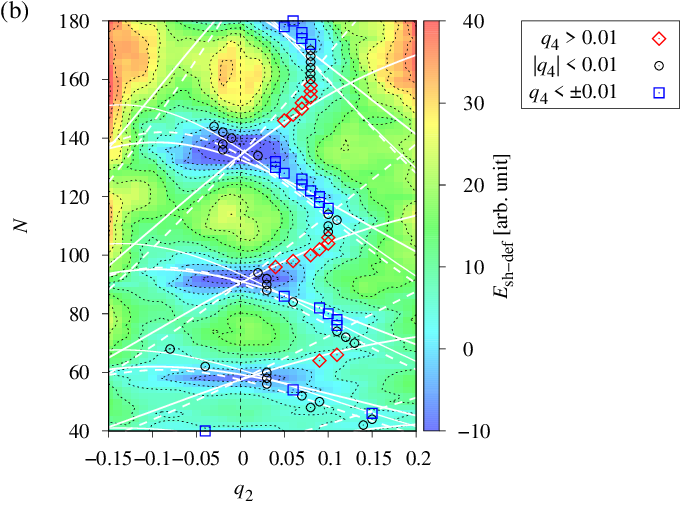} 
\end{flushright}
\caption{\label{fig:vmin_q}
Contour plot of shell-deformation energy for quadrupole parameter
$q_2$ and particle number $N$.  Panel (a) is for pure spheroidal
shape, and circles indicate the ground-state deformation $q_2$
for given $N$.  In panel (b), hexadecapole deformation is taken into
account and determined so as to minimize the energy $E(N)$ for each
$q_2$.  Diamonds, squares and circles indicate the ground-state
deformations with the
hexadecapole parameter $q_4>0.01$, $q_4<-0.01$ and $|q_4|<0.01$,
respectively.  White solid and dashed curves indicate the
constant action curves for major POs (see text for details) with and
without considering hexadecapole deformation, respectively.}
\end{figure*}

On the other hand, no such significant peak enhancement with varying
$q_4$ are found for the oblate case (b).  Dominant orbits are still
meridian triangular (M3) and parallelogram (M4) families, whose
lengths are not very sensitive to $q_4$
and they cannot make a driving force for the hexadecapole deformation.
One possible effect of the hexadecapole to enhance oblate shell effect
is found in equatorial orbits, that encounters bifurcation as shown in
the right panel of Fig.~\ref{fig:bif_sph}.  The large surface
curvature at the equatorial plane due to an oblate deformation is
reduced by adding the $\beta_4<0$ hexadecapole term and a spherical
symmetry is locally restored at certain combination of $\beta_2<0$ and
$\beta_4<0$.  Its effect is not remarkable enough in the cavity model
to realize oblate ground state shape, but their contributions
become more significant in the potential with realistic radial
dependence in which relative magnitudes of equatorial orbits (compared
with the meridian ones) are larger.

As suggested by those results, the contribution of equatorial and
meridian orbits enhance the deformed shell effect on the prolate side,
each with positive and negative hexadecapole parameter.
Let us confirm this effect to the shape evolution by focusing on the
dominant PO contribution to the shell energy.

\subsection{Constant action curves}

Properties of classical POs, such as action and stability,
continuously vary as the functions of the deformation parameter
$\beta$.
Let us assume that the shell energy (\ref{eq:trace_sce}) is dominated
by the contribution of single PO,
\begin{equation}
\delta E(N)=\frac{A_{\rm po}(e_F)}{(T_{\rm po}/\hbar)^2}
\cos\left(\tfrac{1}{\hbar}
S_{\rm po}(e_F)-\tfrac{\pi}{2}\mu_{\rm po}\right).
\label{eq:trace_sce1}
\end{equation}
Then, the energy $E(N,\beta)$ will
show a valley along the constant action curves (CACs) on the
$N$-$\beta$ plane
\begin{equation}
\tfrac{1}{\hbar}S_{\rm po}(e_F(N),\beta)
 -\tfrac{\pi}{2}\mu_{\rm PO}=(2n+1)\pi, \quad \text{($n$: integer)}
\label{eq:e_valley}
\end{equation}
where the cosine function of the PO contribution in (\ref{eq:trace_sce1})
takes the minimum value.
Let us apply this formula to the cavity model.

Due to the short-range nature of nuclear force, the volume of the body
surrounded by the potential surface is usually assumed to be conserved
under deformation and is proportional to the mass number $A$.
Therefore, the radius of the cavity surface in the
spherical limit can be put $R_0(N)=r_0N^{1/3}$ with $r_0$ independent
of $N$.  Then, the action integral along the PO is expressed as
\begin{align}
S_{\rm po}(e_F,\beta)&=\oint_{\rm po}\bp\cdot d\bq
=p_FL_{\rm po}(N,\beta) \NN
&=p_FR_0(N)l_{\rm po}(\beta),
\end{align}
where $l_{\rm po}(\beta)$ is the length of the PO in the deformed
cavity whose volume is same as that of a unit sphere.  The Fermi momentum
$p_F=\hbar k_F$ is known to be approximately independent of $N$ in
nuclei not very far from beta stability line.
Using these properties, equation of the CAC
(\ref{eq:e_valley}) is rewritten as
\begin{gather}
N^{1/3}=\frac{(2n+1+\tfrac12\mu_{\rm po})\pi}{k_Fr_0 l_{\rm po}(\beta)}.
\label{eq:valley_cavity}
\end{gather}

In Fig.~\ref{fig:vmin_q}, the CACs for major POs are shown.
They are drawn overlaid
on the contour plot of the shell-deformation energy\cite{Arita16}
\begin{equation}
E_{\textrm{sh-def}}(N,\beta)\equiv E(N,\beta)-\tilde{E}(N,0),
\label{eq:e_sh_def}
\end{equation}
which evaluates the energy by taking spherical liquid drop as
reference.
Since the lengths of the equatorial orbits decrease with increasing
$q_2$, the CACs become those sloping downward to the right.
The lengths of the meridian orbits are increasing function of $q_2$
in the prolate side, while it is relatively flat in the oblate side.

In the panel (a), shell-deformation energy is calculated by considering
only the quadrupole (spheroidal) shape degree of freedom.  Ground-state
shape
for each $N$ is indicated by a circle symbol.  White broken curves
represent the CACs of the POs (with the Maslov indices evaluated
in the prolate side).  Most of the ground state shapes are
locating along the CACs of meridian orbits in the prolate side,
which explains the predominance of prolate shapes\cite{Frisk90}.
Since the CACs of the meridian orbits are flat
on the oblate side, they cannot produce significant change of gross shell
structure there.  On the other hand, one finds good slope on the prolate
side and nuclei have more chances to gain energy by the deformation.
Looking at the panel (a) of Fig.~\ref{fig:vmin_q}, however, the system
just above the spherical magic number tend to take oblate shapes, and then
turn into prolate shapes with increasing nucleon number.  This
does not match the actual situation of nuclei that take prolate shapes
first and then turn into oblate shapes when the nucleon number increases
from one spherical magic to the next one.

In the panel (b), hexadecapole deformation is additionally considered
and optimized for each value of $q_2$.  White broken curves are the
same for the panel (a) and white solid curves are CACs for POs with
setting $q_4$ at the bifurcation point(s) for each value of $q_2$.
For the CACs of equatorial orbits, $q_4$ is set to the bifurcation point
corresponding to the local spherical symmetry.  For the CACs of meridian
diamond orbit, two downward-right curves correspond to those evaluated
at two successive bifurcation deformations shown in
Fig.~\ref{fig:traj_md}.
The most significant effect of the hexadecapole deformation is the
enhancement of equatorial orbit contribution due to the local
spherical symmetry.  This makes nuclei belonging to
the first halves of the shells to take prolate shape with positive $q_4$.
For nuclei belonging to the second halves of the shells, the
contribution of meridian orbit becomes more significant due to the
effect of codimension-two bifurcation with negative $q_4$.
This scenario clearly and consistently explains the prolate shape
dominance and hexadecapole shape evolution.

\begin{figure}
\centering
\includegraphics[width=\linewidth]{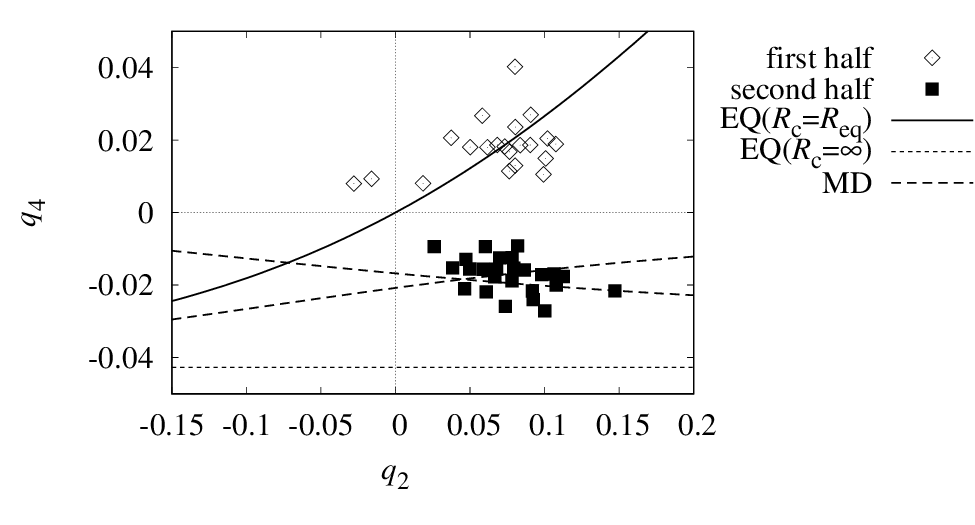} 
\caption{\label{fig:vmin}
Distribution of the ground-state shapes in $(q_2, q_4)$ plane.
Open diamonds and solid squares indicate the results of
the first and second halves of the spherical shell gap between
each magic number and the next, respectively.
Curves represent the bifurcation deformations of major classical
POs (see text for details).}
\end{figure}

For a clearer illustration of the significance of bifurcations, 
correlation between quadrupole and hexadecapole parameters for the
ground-state shapes is
shown in Fig.~\ref{fig:vmin}.  The values of the parameters $(q_2,
q_4)$ are indicated, distinguished by open diamond and solid square
for nuclei whose particle numbers belong to the first and second halves
of the shells.
Solid curve represents the shape where the equatorial orbits encounter
bifurcation due to the restoration of local spherical symmetry.
Most of the diamond symbols distribute along this curve.
Two broken curves represent the shapes where the meridian diamond
orbit encounters two successive pitchfork bifurcations shown in
Fig.~\ref{fig:traj_md}.  The square symbols are concentrated around
the crossing
point $(q_2,q_4)\sim(0.05,-0.02)$ of these two curves, where a
local symmetry with higher dimension is restored (codimension-two bifurcation).
In this way, effect of PO bifurcation taking places at certain combinations of
quadrupole and hexadecapole parameters clearly explains the origin of the
quadrupole-hexadecapole correlation and hexadecapole shape evolution
in ground-state shape evolution.

In a calculation with more realistic mean field potential, oblate ground
states appear near the last part of the shells.  This could be
explained by the bifurcation of equatorial orbits of the
type displayed in the right panel of Fig.~\ref{fig:bif_sph}:
spherical symmetry is locally restored around the equatorial plane
for certain combinations of negative $q_2$ and negative $q_4$.
By considering surface diffuseness of the potential, the relative
magnitudes of the equatorial orbit contributions enhance and
the oblate valley of equatorial orbit CAC surpasses the prolate valley
of the meridian orbit CAC.  This can be
actually examined in the power-law potential model
(\ref{eq:mpot}) putting $\alpha$ sufficiently larger than that of the
oscillator.  The semiclassical analysis of power-law potential model
(\ref{eq:mpot}) with
intermediate values of $\alpha$, also including spin-orbit coupling, is
in preparation.

\section{Summary}
\label{sec:summary}

The mechanism for the systematic evolution of hexadecapole deformation
suggested in realistic theoretical calculations is investigated in
simplified mean-field potential models.  In both cavity and oscillator
type potential models, the same kind of hexadecapole shape evolution
is found.  Semiclassical analysis of the gross shell structure is
carried out for the cavity potential model.  For the systems
belonging to the first halves of the shells, bifurcation of
equatorial orbits that occur at specific combination of prolate and positive
hexadecapole deformation play the essential role in the enhancement of
deformed shell effect.  For the systems
belonging to the second halves of the shells, codimension-two bifurcation of
the meridian diamond orbits that occur at specific
combination of prolate and negative hexadecapole deformation play a
significant role.  The effect of these bifurcations also provide
stronger support for the mechanism of prolate-shape predominance in
nuclear ground states.

When one considers quadrupole shape degree of freedom alone, shell
effect rapidly decreases with increasing deformation.  However, a
suitable combination of two different shape degrees of freedom,
quadrupole and hexadecapole, can partially recover the symmetry and
can keep strong shell effect in deformed states.  This phenomenon
manifest itself differently depending on the system (cavity or
oscillator), but the essential mechanism may be understood in the same
footing.  For instance, the codimension-two bifurcation of diamond
orbit in the cavity model will take another form in a different model,
say, with finite power parameter $\alpha$ in Eq.~(\ref{eq:mpot}), as
the mechanism of enhanced deformed shell effect.  On the other hand,
the restoration of local spherical symmetry takes place in general
systems.  Then, the same kind of bifurcation enhancement will be
expected as the origin of hexadecapole shape evolution found for more
realistic mean-field potential models.  The analysis along this line is
one of the important subjects in our future study.

\bibliographystyle{apsrev4-2}
\bibliography{refs}
\end{document}